\documentclass[journal]{IEEEtran}
\IEEEoverridecommandlockouts

\usepackage{cite}
\usepackage{amsmath,amssymb,amsfonts}
\usepackage{graphicx}
\usepackage{textcomp}
\usepackage{xcolor}
\usepackage{caption, float}
\usepackage{subfig}

\usepackage{amsthm}

\newtheoremstyle{mytheoremstyle}
  {0.5pt}    
  {0.5pt}    
  {\itshape} 
  {}         
  {\bfseries}
  {.}         
  { }   
  {\thmname{#1}\thmnumber{~#2}\thmnote{~(#3)}}     

\theoremstyle{mytheoremstyle}

\newtheorem{theorem}{Theorem}

\newtheorem{remark}{Remark}

\usepackage{bm}
\usepackage{algorithm}
\usepackage{algpseudocode}
\usepackage{setspace}
\usepackage{tikz-cd}
\usepackage{makecell}

\usepackage{enumitem}

\newcommand{\br}{\mathbb{R}}

\allowdisplaybreaks[4]
\usepackage{titlesec}
\titlespacing{\section}{0pt}{*0.5}{*0.5}
\titlespacing{\subsection}{0pt}{*0.5}{*0.5}

\title{\Huge{A Simplified Algorithm for Joint Real-Time Synchronization, NLoS Identification, and Multi-Agent Localization}}

\author{Yili Deng, Jie Fan, Jiguang He, Baojia Luo, Miaomiao Dong, Zhongyi Huang\thanks{The first two authors contributed equally to this work (Corresponding author: Jiguang He).

Y. Deng, J. Fan, and Z. Huang are with the Department of Mathematical Sciences, Tsinghua University, Beijing 100084, China (e-mail: dengyl21@mails.tsinghua.edu.cn, fanj21@mails.tsinghua.edu.cn, zhongyih@tsinghua.edu.cn).

J. He is with Great Bay University, Dongguan 523000, China, Great Bay Institute for Advanced Study (GBIAS), Dongguan 523000, China, and Center for Wireless Communications, University of Oulu, Oulu 90014, Finland (e-mail: jiguang.he@gbu.edu.cn).

B. Luo and M. Dong are with the Theory Lab, Central Research Institute, 2012 Labs, Huawei Technologies Co. Ltd., HONG KONG SAR, China (e-mail: luobaojia2@huawei.com, dong.828599@huawei.com).
}
}

\definecolor{mygray}{gray}{0.6}

\begin{document}
\maketitle

\begin{abstract}
Real-time, high-precision localization in large-scale wireless networks faces two primary challenges: clock offsets caused by network asynchrony and non-line-of-sight (NLoS) conditions. 
To tackle these challenges, we propose a low-complexity real-time algorithm for joint synchronization and NLoS identification-based localization.
For precise synchronization, we resolve clock offsets based on accumulated time-of-arrival measurements from all the past time instances, modeling it as a large-scale linear least squares (LLS) problem.
To alleviate the high computational burden of solving this LLS, we introduce the blockwise recursive Moore-Penrose inverse (BRMP) technique, a generalized recursive least squares approach, and derive a simplified formulation of BRMP tailored specifically for the real-time synchronization problem. Furthermore, we formulate joint NLoS identification and localization as a robust least squares regression (RLSR) problem and address it by using an efficient iterative approach.
Simulations show that the proposed algorithm achieves sub-nanosecond synchronization accuracy and centimeter-level localization precision, while maintaining low computational overhead.


\begin{IEEEkeywords}
Time-of-arrival, localization, synchronization, non-line-of-sight, blockwise recursive Moore-Penrose inverse
\end{IEEEkeywords}

\end{abstract}

\section{Introduction}~\label{sec: intro}
Future large-scale wireless sensor networks (WSNs) are expected to support real-time high-precision device localization, enabling applications such as intelligent transportation, environmental monitoring, and the Internet of Things \cite{shastri2022review}.
A typical large-scale WSN consists of multiple anchors with known positions and multiple agents to be localized.
Time-of-arrival (ToA) measurements are commonly used for localization as they accurately capture the true distances between anchors and agents~\cite{laoudias2018survey}. 
However, practical ToA measurements are often impacted by two major non-ideal factors: clock offsets and non-line-of-sight (NLoS) conditions.

Clock offsets arise due to clock asynchrony between the
agent and each anchor. 
Existing methods typically address this by using multiple calibration emitters with known positions \cite{zou2020semidefinite, ma2019direct} or communication between anchors \cite{wang2020tdoa}.
However, these approaches introduce additional overhead. 
Alternatively, joint localization and synchronization (JLAS) can be achieved solely using ToA measurements between multiple agents and anchors across different time instances.
Techniques such as maximum likelihood (ML) estimation \cite{jean2014passive} and weighted least squares \cite{guo2022new, wang2011robust} have been proposed for this task.
Nevertheless, since these methods rely on ToAs from all time instances to achieve precise synchronization, their computational complexity increases rapidly with time, limiting their applicability for real-time localization in large-scale WSNs.


NLoS conditions are another significant hurdle for high-precision localization, especially in indoor or urban environments.
In NLoS conditions, line-of-sight (LoS) paths are blocked and the ToA measurements can be corrupted by severe NLoS errors \cite{yousefi2015sensor}.
To improve localization accuracy, a common strategy is to discard NLoS ToAs and utilize the remaining LoS ToAs for localization. 
To identify NLoS ToAs, a subspace-based method was proposed in \cite{momtaz2018nlos} using the statistical characteristics of ToA data, and machine learning-based classification models were applied in \cite{huang2020machine}.
However, these methods require extensive data collection, increasing localization overhead.
Additionally, these methods assume that all anchors are synchronized. 
In asynchronous networks, joint NLoS identification and localization becomes a challenging problem that remains unresolved in existing literature.
Furthermore, in large-scale asynchronous WSNs under NLoS conditions, utilizing a large number of ToA measurements across multiple time instances to achieve real-time high-precision localization is even more challenging.

This paper proposes a low-complexity algorithm for real-time synchronization, NLoS identification, and multi-agent localization.
The algorithm consists of two parts: synchronization and NLoS identification-based localization.
At time instance $t$, we model synchronization as a large-scale linear least squares (LLS) problem, using ToA measurements collected over all historical $t$ time instances.
Traditional LLS solutions are computationally intensive as $t$ grows.
To address this, we introduce the blockwise recursive Moore-Penrose inverse (BRMP) technique \cite{zhuang2021blockwise}.
As a generalized recursive least squares (RLS) approach, BRMP transforms the LLS solution into a clock offset update process that only depends on the newly measured ToAs at time instance $t$.
In this way, solving LLS does not involve historical ToAs before time $t$, thus making the computation and memory costs independent of $t$.
We derive a simplified version of the BRMP formula for real-time synchronization, further reducing computational complexity.
Moreover, we formulate the joint NLoS identification and localization as a robust regression-based combinatorial optimization problem, which is then solved using an explicit cross-iteration approach.
Simulations demonstrate that the proposed algorithm achieves clock offset and position estimation errors of less than $0.1$ ns and $10$ cm over time, respectively, while maintaining a low runtime.

\section{System Model and Problem Formulation}~\label{sec: System_Model}
Consider an $l$-dimensional ($l=2$ or $3$) WSN consisting of $M$ fixed anchors indexed by the $\mathcal{M} = \{1, 2, \ldots, M\}$, and $N$ moving agents to be located.
Denote the known coordinates of the $m$-th anchor by $\mathbf{q}_m$ and the unknown coordinates of the $n$-th agent at time instance $t$ by $\mathbf{p}_{t,n}$.
The agent's position $\mathbf{p}_{t,n}$ can either be randomly generated over time $t$ or follow a predefined trajectory.
Each $m$-th anchor has an internal clock offset $\delta_{m}$.
We consider a quasi-synchronous network \cite{wang2011robust} where clock skew is negligible, allowing clock offset $\delta_{m}$ to be assumed constant over time\footnote{For networks with significant clock skew, the clock skew can be estimated and corrected using the least-squares estimator proposed in \cite{li2018joint}.}.

At different time instance $t$, we collect the ToA measurement $r_{t, n, m}$ received by the $m$-th anchor from the $n$-th agent.
The ToA $r_{t, n, m}$ is the time delay of  either a LoS or NLoS path, and is given by
\begin{align}
	r_{t,n, m} = \frac{1}{c} \| \mathbf{q}_m - \mathbf{p}_{t,n} \| + \tau_{t,n} + \delta_{m} + b_{t,n, m} + \epsilon_{t,n, m}, \label{eq: toa}
\end{align}
where $c$ is the speed of radio propagation, $\tau_{t,n}$ is the unknown transmission time at the $n$-th agent,  
$\epsilon_{t,n, m}$ is the measurement noise following a zero-mean Gaussian i.i.d. distribution with the standard deviation (STD) $\sigma$, and
$b_{t,n, m}$ represents the potential NLoS error. 
Specifically, $b_{t,n,m} = 0$ when $m \in \mathcal{S}_{t, n}^{\mathrm{LoS}}$, and $b_{t,n, m} \gg |\epsilon_{t,n, m}|$ when $m \in \mathcal{S}_{t, n}^{\mathrm{NLoS}}$, where 
$\mathcal{S}_{t, n}^{\mathrm{LoS}} \subseteq \mathcal{M}$ and $\mathcal{S}_{t, n}^{\mathrm{NLoS}} = \mathcal{M} - \mathcal{S}_{t, n}^{\mathrm{LoS}}$ denote the index subsets of LoS ToAs and NLoS ToAs from the $n$-th agent at time instance $t$, respectively.

By defining the vectors associated with the $n$-th agent at time instance $t$ as $\mathbf{r}_{t, n} \!=\! [ r_{t,n, 1}, \ldots, r_{t,n, M} ]^{\mathrm{T}}$, $\mathbf{d} \left( \mathbf{p}_{t,n} \right) \!=\! \frac{1}{c} [ \| \mathbf{q}_1 - \mathbf{p}_{t,n} \|, \ldots,  \| \mathbf{q}_M - \mathbf{p}_{t,n} \| ]^{\mathrm{T}}$, $\mathbf{1}_{M} \!=\! [ 1, \ldots, 1 ]^{\mathrm{T}} \in \br^M$, $\boldsymbol{\delta} \!=\! [ \delta_1, \ldots, \delta_M ]^{\mathrm{T}}$, $\mathbf{b}_{t, n} \!=\! [ b_{t,n, 1}, \ldots, b_{t,n, M} ]^{\mathrm{T}}$, and $\boldsymbol{\epsilon}_{t, n} \!=\! [ \epsilon_{t,n, 1}, \epsilon_{t,n, 2}, \ldots, \epsilon_{t,n, M} ]^{\mathrm{T}} $, Eq.~\eqref{eq: toa} is rewritten as
\begin{align}
	\mathbf{r}_{t, n} = \mathbf{d} \left( \mathbf{p}_{t,n} \right) + \tau_{t,n} \mathbf{1}_{M} + \boldsymbol{\delta} + \mathbf{b}_{t, n} + \boldsymbol{\epsilon}_{t, n}. \label{eq: toa_vec}
\end{align}
We aim to select the LoS ToA measurements for JLAS.
In the case where only LoS ToAs from $ \mathcal{S}_{t, n}^{\mathrm{LoS}}$ are considered, Eq.~\eqref{eq: toa_vec} can be expressed as
\begin{align}
	\mathbf{F} \left( {\mathcal{S}_{t, n}^{\mathrm{LoS}}} \right) \left( \mathbf{r}_{t, n} - \mathbf{d} \left( \mathbf{p}_{t,n} \right) - \tau_{t,n} \mathbf{1}_{M} - \boldsymbol{\delta} - \boldsymbol{\epsilon}_{t, n} \right) = \mathbf{0}, \label{eq: toa_vec_los}
\end{align}
where $\mathbf{F}(\mathcal{S}) \in \mathbb{R}^{|\mathcal{S}| \times M}$ is the selection matrix corresponding to the ToA subset $\mathcal{S} \subseteq \mathcal{M}$ with $|\mathcal{S}|$ denoting the cardinality of $\mathcal{S}$.
Given $\mathcal{S} \!=\! \{s_1, s_2, \ldots, s_{|\mathcal{S}|}\}$, $\mathbf{F}(\mathcal{S})$ is defined such that $\left[\mathbf{F} \left( \mathcal{S} \right) \right]_{i, j} \!=\! 1$ if $s_i \!=\! j$, and $\left[\mathbf{F} \left( \mathcal{S} \right) \right]_{i, j} \!=\! 0$ otherwise.
When $\mathbf{F}(\mathcal{S})$ multiplies a vector, it selects the components of the vector corresponding to the indices in $\mathcal{S}$, forming a new vector.

The problem of synchronization, NLoS identification, and multi-agent localization can be defined as: At time $t$, use a total of $tMN$ accumulated ToA measurements $r_{u,n, m}$ to estimate clock offset $\delta_{m}$, LoS ToA subset ${\mathcal{S}_{u, n}^{\mathrm{LoS}}}$, and agent's position $\mathbf{p}_{u,n}$, where $u = 1, \ldots, t$, $n = 1, \ldots, N$, and $m = 1, \ldots, M$. 
We formulate the ML estimator for this problem as
\begin{align*}
	( \mathcal{P}1 )\!: \!\!\!\! &\min_{ \mathbf{p}_{u,n}, \boldsymbol{\delta}, \atop \mathcal{S}_{u,n}, \tau_{u,n} } \! \sum_{u=1}^{t} \! \sum_{n=1}^{N} \big\| \mathbf{F} ( \mathcal{S}_{u,n} ) \!\left( \mathbf{r}_{u,n} \!-\! \mathbf{d} \left( \mathbf{p}_{u,n} \right) \!-\! \tau_{u,n} \mathbf{1}_{M} \!-\! \boldsymbol{\delta} \right) \!\big\|^2  \\
    & \quad \text{s.t.} \quad |\mathcal{S}_{u,n}| \geq \alpha M,\; u = 1, 2, ..., t,\, n = 1, 2, ..., N,
\end{align*}
where $\| \cdot \|$ denotes the Euclidean norm.
Note that $\mathcal{S}_{u,n} = \emptyset$ automatically minimizes the objective function of $( \mathcal{P}1 )$ to $0$.
However, in practical large-scale WSNs, the LoS ToA subset is rarely empty.
To address this, we impose the constraint $|\mathcal{S}_{u,n}| \geq \alpha M$, ensuring that at least $\alpha M$ ToA measurements are LoS-dominated for any time $u$ and agent $n$, where $\alpha$ is a predefined adjustable parameter.
When $\alpha$ is small, problem $( \mathcal{P}1 )$ becomes underdetermined, resulting in non-unique and false solutions. 
To ensure the problem is well-posed, we set $\alpha > \frac{1}{2}$ \cite{bhatia2015robust}.


Furthermore, we aim to solve problem $( \mathcal{P}1 )$ in real-time, i.e., instantly updating clock offset $\boldsymbol{\delta}$ and generating localization result $\mathbf{p}_{t,n}$ as time $t$ progresses.
However, $( \mathcal{P}1 )$ is a complex combinatorial optimization problem since subset ${\mathcal{S}_{u, n}^{\mathrm{LoS}}}$ is unknown. It is challenging to obtain an optimal solution.
Additionally, as $t$ increases (e.g., $t > 1000$), solving $( \mathcal{P}1 )$ based on the total $tMN$ ToA measurements demands significant computational and memory resources. 
Using only a subset of the ToAs helps mitigate this issue. However, the reduction in data leads to lower synchronization and localization accuracy.

\section{Clock Synchronization and Localization}~\label{sec: Clock Synchronization and Localization}
In this section, we propose a novel real-time synchronization and localization algorithm in mixed LoS and NLoS conditions.
This algorithm utilizes all $tMN$ ToA measurements while ensuring that the computational and memory overhead remains independent of time $t$. 
We decompose the original problem $( \mathcal{P}1 )$ into two subproblems: real-time synchronization and NLoS identification-based multi-agent localization, and solve these subproblems sequentially at each time instance $t$.
Subsections \ref{subsec: real-time Synchronization} and \ref{subsec: Sensor Selection and Localization} introduce the respective solutions for these two subproblems.
Subsection \ref{subsec: Algorithm Summary} summarizes the proposed algorithm.

\subsection{Real-Time Synchronization}~\label{subsec: real-time Synchronization}
This subsection demonstrates how to update the estimated clock offset $\hat{\boldsymbol{\delta}}_t$ over time $t$ to achieve real-time synchronization. 
Given the estimated LoS ToA set $\hat{\mathcal{S}}_{u,n}$ and position $\hat{\mathbf{p}}_{u,n}$ for each agent $n$ up to time $t$ (the selection of $\hat{\mathcal{S}}_{u,n}$ and estimation of $\hat{\mathbf{p}}_{u,n}$ will be discussed in Subsection \ref{subsec: Sensor Selection and Localization}), the subproblem for the ML estimator of  $\hat{\boldsymbol{\delta}}_t$ is formulated as
\begin{align*}
	( \mathcal{P}2 )\!: \min_{\boldsymbol{\delta}, \tau_{u,n} } \! \sum_{u=1}^{t}  \sum_{n=1}^{N} \big\| \mathbf{F} ( \hat{\mathcal{S}}_{u,n} ) \!\left( \mathbf{r}_{u,n} \!-\! \mathbf{d} \left( \hat{\mathbf{p}}_{u,n} \right) \!-\! \tau_{u,n} \mathbf{1}_{M} \!-\! \boldsymbol{\delta} \right) \!\big\|^2. 
\end{align*}
Note that the ML cost function is minimized by $\tau_{u,n} = \frac{1}{|\hat{\mathcal{S}}_{u,n}|}  \mathbf{1}_{|\hat{\mathcal{S}}_{u,n}|}^{\mathrm{T}} \mathbf{F} ( \hat{\mathcal{S}}_{u,n} ) \left( \mathbf{r}_{u,n} - \mathbf{d} \left( \hat{\mathbf{p}}_{u,n} \right) - \boldsymbol{\delta} \right)$.
Substituting this into $( \mathcal{P}2 )$ yields
\begin{align*}
	( \mathcal{P}3 )\!: \min_{\boldsymbol{\delta}} \sum_{u=1}^{t}  \sum_{n=1}^{N} \big\| \mathbf{B} ( \hat{\mathcal{S}}_{u,n} ) \mathbf{F} ( \hat{\mathcal{S}}_{u,n} ) \left( \mathbf{r}_{u,n} \!-\! \mathbf{d} \left( \hat{\mathbf{p}}_{u,n} \right) \!-\! \boldsymbol{\delta} \right) \!\big\|^2, 
\end{align*}
where the matrix $\mathbf{B} ( \hat{\mathcal{S}}_{u,n} )$ is defined as $\mathbf{B} \left( \mathcal{S} \right) = \mathbf{I} - \frac{1}{|\mathcal{S}|} \mathbf{1}_{|\mathcal{S}|} \mathbf{1}_{|\mathcal{S}|}^{\mathrm{T}} \in \br^{|\mathcal{S}| \times |\mathcal{S}|}$ with $\mathbf{I}$ being the identity matrix.

Problem $( \mathcal{P}3 )$ is a large-scale LLS problem. 
By defining $\mathbf{y}_{u, n} = \mathbf{B} ( \hat{\mathcal{S}}_{u,n} ) \mathbf{F} ( \hat{\mathcal{S}}_{u,n} ) \left( \mathbf{r}_{u,n} - \mathbf{d} \left( \hat{\mathbf{p}}_{u,n} \right) \right) \in \br^{|\hat{\mathcal{S}}_{u,n}|}$, 
$\mathbf{y}_{u}= [ \mathbf{y}_{u, 1}^{\mathrm{T}} , \mathbf{y}_{u, 2}^{\mathrm{T}}, \cdots,\mathbf{y}_{u, N}^{\mathrm{T}} ]^{\mathrm{T}} \in \br^{M_u}$, 
$\mathbf{A}_{u, n} = \mathbf{B} ( \hat{\mathcal{S}}_{u,n} ) \mathbf{F} ( \hat{\mathcal{S}}_{u,n} ) \in \br^{|\hat{\mathcal{S}}_{u,n}| \times M}$, and 
$\mathbf{A}_{u} = [ \mathbf{A}_{u, 1}^{\mathrm{T}}, \cdots, \mathbf{A}_{u, N}^{\mathrm{T}} ]^{\mathrm{T}} \in \br^{M_u \times M}$ with $M_u = \sum_{n=1}^{N} |\hat{\mathcal{S}}_{u,n}|$, LLS $( \mathcal{P}3 )$ can be expressed in a compact form as
\begin{align*}
    ( \mathcal{P}4 )\!:  \min_{\boldsymbol{\delta}} \sum_{u=1}^{t}  \big\| \mathbf{y}_{u} - \mathbf{A}_{u} \boldsymbol{\delta} \big\|^2.
\end{align*}
More generally, we consider a weighted LLS problem
\begin{align*}
    ( \mathcal{P}5 )\!: \min_{\boldsymbol{\delta}} \sum_{u=1}^{t} \lambda^{t-u} \big\| \mathbf{y}_{u} - \mathbf{A}_{u} \boldsymbol{\delta} \big\|^2, 
\end{align*}
where $0 < \lambda \leq 1$ is a forgetting factor. The parameter $\lambda$ governs the influence of past ToA measurements by exponentially decreasing their weight as they age.
This allows the model to prioritize more recent data and adapt to changes over time \cite{islam2019recursive}.
$( \mathcal{P}4 )$ is a special case of $( \mathcal{P}5 )$ when $\lambda = 1$.

It is easy to verify that $\sum_{i=1}^{|\hat{\mathcal{S}}_{u,n}|} \left[ \mathbf{A}_{u, n} \right]_{i, :} = \mathbf{0}$, indicating that $\mathbf{A}_t$ is rank-deficient. 
Traditional methods for solving LLS problems use the Moore-Penrose (MP) inverse to provide an exact solution of $( \mathcal{P}5 )$, denoted as $\hat{\boldsymbol{\delta}}_t$.
However, the computational complexity of the MP inverse for solving $( \mathcal{P}5 )$ is $O\left( \big(\sum_{u=1}^{t} M_u\big)^3 \right)$ \cite{ben2003generalized}.
Clearly, directly using the MP inverse incurs substantial memory usage and computational overhead as time $t$ increases.
This is not conducive to low-complexity real-time synchronization in large-scale WSNs.

To reduce computational resource consumption, we exploit the block-based partitioning structure of LLS $( \mathcal{P}5 )$ and apply the BRMP technique to update clock offset $\hat{\boldsymbol{\delta}}_t$ in real-time.
BRMP is a state-of-the-art RLS technique that accommodates rank-deficient matrices \cite{zhuang2021blockwise}.
Instead of recomputing the LLS solution from scratch with each new measurement, RLS updates the solution incrementally,  significantly reducing the computational complexity. 
To the best of our knowledge, this paper is the first to introduce RLS-related techniques into the field of JLAS.
Theorem \ref{theorem 0} presents the BRMP formula to solve~$( \mathcal{P}5 )$.

\begin{theorem}[BRMP formula]
    Using the clock offset's solution $\hat{\boldsymbol{\delta}}_{t-1}$ from the previous time instance, the new block of coefficient matrix $\mathbf{A}_{t}$, and the new block $\mathbf{y}_{t}$, BRMP provides the exact solution of $( \mathcal{P}5 )$ through recursive formula
    \begin{align}
    	\hat{\boldsymbol{\delta}}_t&= \hat{\boldsymbol{\delta}}_{t-1} + \mathbf{G}_t ( \mathbf{y}_t-\mathbf{A}_t \hat{\boldsymbol{\delta}}_{t-1} ) \in \br^{M} , \label{eq: RLS_delta}
    \end{align}
    where recursive formula for $\mathbf{G}_t$ is given by
    \begin{subequations}
    \begin{align}
    &\mathbf{C}_t\!=\! \mathbf{A}_t\mathbf{Q}_{t-1} \in \br^{M_t \!\times\! M} , \label{eq: RLS_C} \\
     &\mathbf{K}_t\!=\! (\mathbf{I} \!+\! (\mathbf{I}-\mathbf{C}_t \mathbf{C}_t^{\dagger})\mathbf{A}_t\mathbf{R}_{t-1}\mathbf{A}_t^{\mathrm{T}} (\mathbf{I} \!-\! \mathbf{C}_t \mathbf{C}_t^{\dagger})) ^{-1} \in \br^{M_t \!\times\! M_t} , \label{eq: RLS_K} \\
     &\mathbf{V}_t\!=\! (\mathbf{I}-\mathbf{C}_t^{\dagger}\mathbf{A}_t)\mathbf{R}_{t-1}\mathbf{A}_t^{\mathrm{T}} \mathbf{K}_t(\mathbf{I}-\mathbf{C}_t \mathbf{C}_t^{\dagger}) \in \br^{M \!\times\! M_t},\\
     &\mathbf{G}_t\!=\! \mathbf{C}_t^{\dagger}+\mathbf{V}_t \in \br^{M \!\times\! M_t} ,\\
     &\mathbf{Q}_t\!=\! \mathbf{Q}_{t-1}-\mathbf{G}_t\mathbf{A}_t\mathbf{Q}_{t-1} \in \br^{M \!\times\! M} ,\\
     &\mathbf{R}_t\!=\! \frac{1}{\lambda^2} [ (\mathbf{I} \!-\! \mathbf{G}_t\mathbf{A}_t)\mathbf{R}_{t\!-\!1}(\mathbf{I} \!-\! \mathbf{G}_t\mathbf{A}_t)^{\mathrm{T}} \!\!\!+\! \mathbf{G}_t\mathbf{G}_t^{\mathrm{T}} ] \!\in\! \br^{M \!\times\! M} \!\!\!,
    \end{align}
    \label{eq: RLS}
    \end{subequations}
    \!\!where $(\cdot)^{\dagger}$ represents the MP inverse.
    Notably, $M_t =$ $ \sum_{n} |\hat{\mathcal{S}}_{t,n}|$ $ < NM$, indicating that the sizes of matrices $\mathbf{C}_t, \mathbf{K}_t, \mathbf{V}_t,$ $ \mathbf{G}_t, \mathbf{Q}_t, \mathbf{R}_t$ are bounded by a time-independent upper limit.
    The initialization for \eqref{eq: RLS} is $\hat{\boldsymbol{\delta}}_0= \mathbf{0}, \mathbf{Q}_0=\mathbf{I}, \mathbf{R}_0=\mathbf{0}$.
    \label{theorem 0}
\end{theorem}
\begin{IEEEproof}
    By defining the vector
    $
        \tilde{\mathbf{y}}_{t} = [ \frac{1}{\lambda} \mathbf{y}_{1}^{\mathrm{T}}, \frac{1}{\lambda^{2}} \mathbf{y}_{2}^{\mathrm{T}}, \cdots, $ $ \frac{1}{\lambda^{t}} \mathbf{y}_{t}^{\mathrm{T}} ]^{\mathrm{T}} 
        \in \br^{\sum_{u=1}^{t} M_u} 
    $
    and the coefficient matrix 
    $
        \tilde{\mathbf{A}}_{t} \!=\! [ \frac{1}{\lambda} \mathbf{A}_{1}^{\mathrm{T}}, $ $ \frac{1}{\lambda^{2}} \mathbf{A}_{2}^{\mathrm{T}}, \cdots, \frac{1}{\lambda^{t}} \mathbf{A}_{t}^{\mathrm{T}} ]^{\mathrm{T}}
        \in \br^{\left( \sum_{u=1}^{t} M_u \right)  \times M}, 
    $
    LLS $( \mathcal{P}5 )$ is simplified to
    \begin{align*}
        ( \mathcal{P}6 )\!: \min_{\boldsymbol{\delta}} \big\| \tilde{\mathbf{y}}_{t} - \tilde{\mathbf{A}}_{t} \boldsymbol{\delta} \big\|^2. 
    \end{align*}
    We use MP inverse to provide an exact solution of $( \mathcal{P}6 )$ as 
    \begin{align}
    	\hat{\boldsymbol{\delta}}_t = \tilde{\mathbf{A}}_{t}^{\dagger} \tilde{\mathbf{y}}_{t}. \label{eq: Moore-Penrose}
    \end{align}
    According to the BRMP equations (20)-(26) in \cite{zhuang2021blockwise}, $\hat{\boldsymbol{\delta}}_t$ has an equivalent iterative form as shown in \eqref{eq: RLS_delta}-\eqref{eq: RLS}.
\end{IEEEproof}

Since formulas \eqref{eq: RLS_delta} and \eqref{eq: RLS} only involve the new block $\mathbf{A}_{t} \in \br^{M_t  \times M}$ and $\mathbf{y}_{t} \in \br^{ M_t }$ at each time $t$, rather than the entire matrix $\tilde{\mathbf{A}}_{t} \in \br^{\left( \sum_{u=1}^{t} M_u \right)  \times M} $ and the entire $\tilde{\mathbf{y}}_{t} \in \br^{\sum_{u=1}^{t} M_u}$, the memory usage and computational overhead of BRMP are independent of time $t$, making it well-suited for real-time synchronization.

Furthermore, we demonstrate that the recursive formula \eqref{eq: RLS} can be simplified when the conditions of the following Theorem \ref{theorem 1} are satisfied.
Define the total selected ToA set from time instance $1$ to $t$ as $\tilde{\mathcal{S}}_{t} = \bigcup_{u=1}^{t} \bigcup_{n=1}^{N} \hat{\mathcal{S}}_{u,n}$.

\begin{theorem}[Simplified BRMP formula]
    Assume that the number of ToAs in $\hat{\mathcal{S}}_{u,n}$ is always greater than half of the anchor count, i.e., $|\hat{\mathcal{S}}_{u,n}| > \frac{M}{2}, \, \forall u, n$. 
    If the set $\hat{\mathcal{S}}_{t,n}$ satisfies $\bigcup_{n=1}^{N} \hat{\mathcal{S}}_{t,n} \subseteq \tilde{\mathcal{S}}_{t-1}$, then $\mathbf{C}_t = \mathbf{0}$ in \eqref{eq: RLS_C} and the recursive formulas \eqref{eq: RLS} are simplified as
    \begin{subequations}
    \begin{flalign}
        &\mathbf{K}_t \!=\! (\mathbf{I}+\mathbf{A}_t\mathbf{R}_{t-1}\mathbf{A}_t^{\mathrm{T}} ) ^{-1} \in \br^{M_t \!\times\! M_t}, \label{eq:RLS_reduced_K} \\
    	&\mathbf{G}_t \!=\! \mathbf{R}_{t-1} \mathbf{A}_t^{\mathrm{T}} \mathbf{K}_t \in \br^{M \!\times\! M_t},\\
        &\mathbf{Q}_t \!=\! \mathbf{Q}_{t-1}-\mathbf{G}_t\mathbf{A}_t\mathbf{Q}_{t-1} \in \br^{M \!\times\! M},\\
        &\mathbf{R}_t\!=\! \frac{1}{\lambda^2} [ (\mathbf{I} \!-\! \mathbf{G}_t\mathbf{A}_t)\mathbf{R}_{t\!-\!1}(\mathbf{I} \!-\! \mathbf{G}_t\mathbf{A}_t)^{\mathrm{T}} \!\!\!+\! \mathbf{G}_t\mathbf{G}_t^{\mathrm{T}} ] \!\in\! \br^{\! M \!\times\! M} \!\!\!.
    \end{flalign}
    \label{eq:RLS_reduced}
    \end{subequations}
\label{theorem 1}   
\end{theorem}
\begin{IEEEproof}
    Define $\mathbf{g}_{i, j} \in \br^M$ ($i \neq j$) as a vector with the $i$-th element equal to $1$, the $j$-th element equal to $-1$, and all other elements equal to $0$. 
    For any set $\mathcal{S} = \{s_1, s_2, \ldots, s_{|\mathcal{S}|}\} \subseteq \mathcal{M}$, define the set 
    $$\mathcal{D} \left( \mathcal{S} \right) = \{ \mathbf{g}_{i, j} : i \in \mathcal{S}, j \in \mathcal{S}, i \neq j \},$$ 
    and define the matrix $\mathbf{A}\left( \mathcal{S} \right) = \mathbf{B} \left( \mathcal{S} \right) \mathbf{F} \left( \mathcal{S} \right)$. 
    Then, $\mathbf{A}_{u, n} = \mathbf{A}( \hat{\mathcal{S}}_{u,n} ) $.
    Let $\mathrm{span} \left( \mathcal{D} \right)$ denote the linear space spanned by vectors in the set $\mathcal{D}$, and $\mathrm{rowsp}(\mathbf{A})$ denote the row space of matrix $\mathbf{A}$.

    We first prove that for any set $\mathcal{S} \subseteq \mathcal{M}$, 
    $\mathrm{span} \left( \mathcal{D} \left( \mathcal{S} \right) \right) = \mathrm{rowsp}(\mathbf{A}\left( \mathcal{S} \right))$
    always holds. 
    On one hand, for any $s_i, s_j \in \mathcal{S}$ ($i \neq j$), we have
    $$\mathbf{g}_{s_i, s_j} = [\mathbf{A}\left( \mathcal{S} \right)]_{i, :} - [\mathbf{A}\left( \mathcal{S} \right)]_{j, :}.$$ 
    This implies that 
    \begin{align}
        \mathcal{D} \left( \mathcal{S} \right)  \subseteq \mathrm{rowsp}(\mathbf{A}\left( \mathcal{S} \right)).\label{eq: Theorem 2 1}
    \end{align}
    On the other hand, we observe that
    $$[\mathbf{A}\left( \mathcal{S} \right)]_{i, :} = \frac{1}{|\mathcal{S}|} \sum_{j=1, j \neq i}^{M} \mathbf{g}_{s_i, s_j},$$ 
    which shows that   
    \begin{align}
        \mathrm{rowsp}(\mathbf{A}\left( \mathcal{S} \right)) \subseteq \mathrm{span} \left( \mathcal{D} \left( \mathcal{S} \right) \right). \label{eq: Theorem 2 2}
    \end{align}
    Based on \eqref{eq: Theorem 2 1} and \eqref{eq: Theorem 2 2}, we conclude that  
    $$\mathrm{span} \left( \mathcal{D} \left( \mathcal{S} \right) \right) = \mathrm{rowsp}(\mathbf{A}\left( \mathcal{S} \right)).$$

    Next, we prove that for any $\mathcal{S}_1, \mathcal{S}_2 \subseteq \mathcal{M}$ satisfing $|\mathcal{S}_1| > \frac{M}{2}, |\mathcal{S}_2| > \frac{M}{2}$, it holds that $\mathrm{span} \left( \mathcal{D} \left( \mathcal{S}_1 \bigcup \mathcal{S}_2 \right) \right) = \mathrm{span} \left( \mathcal{D} \left( \mathcal{S}_1 \right) \bigcup \mathcal{D} \left( \mathcal{S}_2 \right) \right).$ 
    According to the definition of $\mathcal{D}(\cdot)$, it is obvious that
    \begin{align}
        \mathcal{D} \left( \mathcal{S}_1 \right) \bigcup \mathcal{D} \left( \mathcal{S}_2 \right) \subseteq \mathrm{span} \left( \mathcal{D} \left( \mathcal{S}_1 \bigcup \mathcal{S}_2 \right) \right). \label{eq: Theorem 2 3}
    \end{align}
    Since $|\mathcal{S}_1| > \frac{M}{2}$ and $|\mathcal{S}_2| > \frac{M}{2}$, we have $\mathcal{S}_1 \cap \mathcal{S}_2 \neq \emptyset$.
    Let $k \in \mathcal{S}_1 \cap \mathcal{S}_2$.
    For any $i \in \mathcal{S}_1, j \in \mathcal{S}_2$, $\mathbf{g}_{i, j} = \mathbf{g}_{i, k} + \mathbf{g}_{k, j}$, hence 
    \begin{align}
        \mathcal{D} \left( \mathcal{S}_1 \bigcup \mathcal{S}_2 \right) \subseteq \mathrm{span} \left( \mathcal{D} \left( \mathcal{S}_1 \right) \bigcup \mathcal{D} \left( \mathcal{S}_2 \right) \right). \label{eq: Theorem 2 4}
    \end{align}
     Putting together \eqref{eq: Theorem 2 3} and \eqref{eq: Theorem 2 4} yields that
    $$\mathrm{span} \left( \mathcal{D} \left( \mathcal{S}_1 \bigcup \mathcal{S}_2 \right) \right) = \mathrm{span} \left( \mathcal{D} \left( \mathcal{S}_1 \right) \bigcup \mathcal{D} \left( \mathcal{S}_2 \right) \right).$$
    
    Finally, rewrite the row space of \( \mathbf{A}_t \) as
    \begin{small}
        \begin{align*}
        \mathrm{rowsp}(\mathbf{A}_t) 
        &= \mathrm{span} \left( \bigcup_{n=1}^{N} \mathrm{rowsp} \left( \mathbf{A}_{t,n} \right) \right) \\
        &= \mathrm{span} \left( \bigcup_{n=1}^{N} \mathrm{span} \left( \mathcal{D}\left(\hat{\mathcal{S}}_{t,n}\right) \right) \right) \\ 
        &= \mathrm{span} \left( \mathcal{D}\left(\bigcup_{n=1}^{N} \hat{\mathcal{S}}_{t,n}\right) \right). 
        \end{align*}
    \end{small}
    Similarly, we rewrite the row space of \( \tilde{\mathbf{A}}_{t-1} \) as
        \begin{small}
        \begin{align*}
         \mathrm{rowsp}(\tilde{\mathbf{A}}_{t-1}) 
        &= \mathrm{span} \left( \bigcup_{u=1}^{t-1} \bigcup_{n=1}^{N} \mathrm{rowsp} \left( \mathbf{A}_{u,n} \right) \right) \\
        &= \mathrm{span} \left( \bigcup_{u=1}^{t-1} \bigcup_{n=1}^{N} \mathcal{D}\left( \hat{\mathcal{S}}_{u,n}\right) \right) \\
        & =\mathrm{span} \left( \mathcal{D}\left(\bigcup_{u=1}^{t-1} \bigcup_{n=1}^{N} \hat{\mathcal{S}}_{u,n}\right) \right)\\
        &=\mathrm{span} \left( \mathcal{D}\left(\tilde{\mathcal{S}}_{t-1}\right) \right).
        \end{align*}
    \end{small}
    Since \( \bigcup_{n=1}^{N} \hat{\mathcal{S}}_{t,n} \subseteq \tilde{\mathcal{S}}_{t-1} \), we have
    $$
    \mathrm{rowsp}(\mathbf{A}_t) \subseteq \mathrm{rowsp}(\tilde{\mathbf{A}}_{t-1}).
    $$

    According to Lemma 5 in \cite{zhuang2021blockwise}, we have $\mathbf{C}_t = \mathbf{0}$, and then the formulas in \eqref{eq: RLS} are reduced to \eqref{eq:RLS_reduced}.
\end{IEEEproof}

As shown in Theorem~\ref{theorem 1}, under the assumption that $|\hat{\mathcal{S}}_{u,n}| > \frac{M}{2}, \, \forall u, n$, if the estimated LoS ToA set $\hat{\mathcal{S}}_{t,n}$ for each agent $n$ at time $t$ is contained within $\tilde{\mathcal{S}}_{t-1}$, then we can use the reduced recursive formula \eqref{eq:RLS_reduced} instead of \eqref{eq: RLS} to update $\hat{\boldsymbol{\delta}}_t$.
This further reduces computational resources.
Notably, the setting $\alpha > \frac{1}{2}$ ensures that the assumption $|\hat{\mathcal{S}}_{u,n}| > \frac{M}{2}$ always holds.

\subsection{NLoS Identification-Based Multi-Agent Localization}~\label{subsec: Sensor Selection and Localization}
In this subsection, we select the LoS ToA set $\hat{\mathcal{S}}_{u,n}$ and estimate the position $\hat{\mathbf{p}}_{u,n}$ for each agent $n$ as time $t$ progresses.
Given the estimated clock offset $\hat{\boldsymbol{\delta}}_{t}$ from the previous time instance $t$, according to $( \mathcal{P}1 )$, 
we find $\hat{\mathcal{S}}_{u,n}$ and $\hat{\mathbf{p}}_{u,n}$ by formulating the localization subproblem for the $n$-th agent at time $t+1$ as
\begin{align*}
	& ( \mathcal{P}7 )\!: \!\!\!\!\!\! \min\limits_{_{\mathbf{p}, \mathcal{S}, \tau_{t+1,n}, \atop |\mathcal{S}| \geq \alpha M }} \!\!\!\!\! \| \mathbf{F} \left( \mathcal{S} \right) ( \mathbf{r}_{t+1,n} \!-\! \mathbf{d} \left( \mathbf{p} \right) \!-\! \tau_{t+1,n} \mathbf{1}_{M} \!-\! \hat{\boldsymbol{\delta}}_{t} ) \|^2,  \\
	\overset{\text{(a1)}}{\Leftrightarrow} \, & ( \mathcal{P}8 )\!: \!\! \min\limits_{_{\mathbf{p}, \mathcal{S}, \atop |\mathcal{S}| \geq \alpha M }} \| \mathbf{B} \left( \mathcal{S} \right) \mathbf{F} \left( \mathcal{S} \right) ( \mathbf{r}_{t+1,n} - \mathbf{d} \left( \mathbf{p} \right) - \hat{\boldsymbol{\delta}}_{t} ) \|^2, 
\end{align*}
where $\mathcal{S} \subseteq \mathcal{M}$ represents the selected subset of LoS ToAs.
The derivation of (a1) is similar to that of $( \mathcal{P}2 )$-$( \mathcal{P}3 )$.

Problem $( \mathcal{P}7 )$ is equivalent to 
\begin{align}
	\min_{\mathbf{p}, \mathcal{S}, \tau_{t+1,n} \atop |\mathcal{S}| \geq \alpha M} \sum_{m \in \mathcal{S}} \left( r_{t+1,n, m} - \frac{1}{c} \| \mathbf{q}_m - \mathbf{p} \| - \tau_{t+1,n} - [ \hat{\boldsymbol{\delta}}_{t} ]_m  \right)^2, \notag
\end{align}
where the notation $\left[ \boldsymbol{\delta} \right]_m$ denotes the $m$-th element of vector $\boldsymbol{\delta}$.
This is a nonlinear robust least squares regression (RLSR) problem. 
Inspired by the solution for linear RLSR in \cite{bhatia2015robust},
we employ an iterative approach to alternately find the position $\mathbf{p}$ and select the set $\mathcal{S}$. In the $k$-th iteration, 
\begin{subequations}
	\label{eq: iter}
\begin{align}
	\hat{\mathbf{p}}^{(k)} &\!=\! \mathop{\text{argmin}}_{\mathbf{p}} \! \| \mathbf{B} ( \hat{\mathcal{S}}^{(k-1)} ) \mathbf{F} ( \hat{\mathcal{S}}^{(k-1)} ) ( \mathbf{r}_{t+1,n} \!-\! \mathbf{d} \left( \mathbf{p} \right) \!-\! \hat{\boldsymbol{\delta}}_{t} ) \|^2, \label{eq: iter_p} \\
	\mathbf{e}^{(k)} &\!=\! \mathbf{B} \left( \mathcal{M} \right) \left( \mathbf{r}_{t+1,n} - \mathbf{d} ( \hat{\mathbf{p}}^{(k)} )- \hat{\boldsymbol{\delta}}_{t} \right), \label{eq: iter_e} \\
	\hat{\mathcal{S}}^{(k)} &\!=\! \left\{ i \in \{1, 2, \ldots, M\} \; | \; \sigma_k^{-1}(i) \leq \alpha M \right\}. \label{eq: iter_S} 
\end{align}
\end{subequations}
where $\sigma_k(i)$ is the permutation that arranges elements of $\mathbf{e}^{(k)}$ in ascending order by absolute values, i.e., $|[\mathbf{e}^{(k)}]_{\sigma(1)}| \leq |[\mathbf{e}^{(k)}]_{\sigma(2)}| \leq \ldots \leq |[\mathbf{e}^{(k)}]_{\sigma(M)}|$.
Step \eqref{eq: iter_p} returns the agent position $\hat{\mathbf{p}}^{(k)}$; step \eqref{eq: iter_e} computes the residual vector $\mathbf{e}^{(k)}$ based on $( \mathcal{P}8 )$, and the idea of step \eqref{eq: iter_S} is to sort all elements of $\mathbf{e}^{(k)}$ in ascending order and select the indices of the $\alpha M$ smallest elements to form the ToA set $\hat{\mathcal{S}}^{(k)}$.

The initialization for \eqref{eq: iter} is $\hat{\mathcal{S}}^{(0)} = \mathcal{M}$.
The iteration \eqref{eq: iter} terminates when either $\hat{\mathcal{S}}^{(k)} = \hat{\mathcal{S}}^{(k-1)}$ or the iteration count $k$ surpasses the predefined maximum limit $K_\mathrm{max}$.
The estimated LoS ToA set and position at time $t+1$ are set to $\hat{\mathcal{S}}_{t+1,n} = \hat{\mathcal{S}}^{(k)}$ and $\hat{\mathbf{p}}_{t+1,n} = \hat{\mathbf{p}}^{(k)}$ when \eqref{eq: iter} terminates.

\subsection{Algorithm Summary}~\label{subsec: Algorithm Summary}
The real-time algorithm for NLoS identification, localization, and synchronization is summarized in Algorithm~\ref{alg}.
At each new time instance $t$, we perform this algorithm to update the clock offset $\hat{\boldsymbol{\delta}}_{t}$ and estimate multi-agent positions
$\{ \hat{\mathbf{p}}_{t,n} \}_{n = 1}^N$ in real-time.

The computational complexity of Algorithm~\ref{alg} is analyzed as follows: For the NLoS identification-based localization (lines 1-10), the primary complexity bottleneck lies in solving problem \eqref{eq: iter_p}. This problem can be efficiently addressed using the quasi-Newton method \cite{shanno1970conditioning} with a complexity of at most $O(K_\mathrm{ne} M^2)$, where $K_\mathrm{ne}$ is the maximum iteration number for the quasi-Newton method.
For synchronization (lines 11-18), the main complexity arises from matrix inversion operation in \eqref{eq: RLS_K} or \eqref{eq:RLS_reduced_K}, with a maximum complexity of $O((NM)^3)$.
Hence, the overall complexity of Algorithm~\ref{alg} is $O(\max(N^3M^3, K_\mathrm{max} K_\mathrm{ne} N M^2)$.
It is observed that due to the use of BRMP (lines 12-18), the update of $\hat{\boldsymbol{\delta}}_{t}$ leverages information from all past time instances without requiring the storage of data prior to time $t-1$, ensuring computational complexity remains independent of $t$.
\begin{algorithm}[t]
	\renewcommand{\algorithmicrequire}{\textbf{Input:}}
	\renewcommand{\algorithmicensure}{\textbf{Output:}}
		\caption{Real-time Algorithm at Time Instance $t$}
		\label{alg}
    \begin{algorithmic}[1]
        \Require Anchor positions $\{ \mathbf{q}_m \}_{m = 1}^M$, ToA measurements $\{\mathbf{r}_{t,n}\}_{n = 1}^N$ at time $t$, forgetting factor $\lambda$, and proportion parameter $\alpha > \frac{1}{2}$
        \Ensure Estimates of multi-agent positions $\{ \hat{\mathbf{p}}_{t,n} \}_{n = 1}^N$ and clock offset $\hat{\boldsymbol{\delta}}_{t}$ at time $t$
        \For{$n = 1, 2, \cdots, N$}
        \State $\hat{\mathcal{S}}^{(0)} = \mathcal{M}$;
        \For{$k = 1, 2, \cdots, K_\mathrm{max}$}
        \State Update $\hat{\mathbf{p}}^{(k)}$ and select $\hat{\mathcal{S}}^{(k)}$ through \eqref{eq: iter};
        \If{$\hat{\mathcal{S}}^{(k)} = \hat{\mathcal{S}}^{(k-1)}$}
        \State \textbf{break};
        \EndIf
        \EndFor
        \State $\hat{\mathcal{S}}_{t,n} = \hat{\mathcal{S}}^{(k)}$, $\hat{\mathbf{p}}_{t,n} = \hat{\mathbf{p}}^{(k)}$;
        \EndFor
        \State Calculate $\mathbf{A}_t$ and $\mathbf{y}_t$ using ToA measurements $\{\mathbf{r}_{t,n}\}$;
        \If{$\bigcup_{n=1}^{N} \hat{\mathcal{S}}_{t,n} \subseteq \tilde{\mathcal{S}}_{t-1}$}
        \State Update clock offset  $\hat{\boldsymbol{\delta}}^{(t)}$ through \eqref{eq: RLS_delta} and \eqref{eq:RLS_reduced};
        \State $\tilde{\mathcal{S}}_{t} = \tilde{\mathcal{S}}_{t-1}$;
        \Else
        \State Update clock offset  $\hat{\boldsymbol{\delta}}^{(t)}$ through \eqref{eq: RLS_delta}-\eqref{eq: RLS};
        \State $\tilde{\mathcal{S}}_{t} = \tilde{\mathcal{S}}_{t-1} \bigcup (\bigcup_{n=1}^{N} \hat{\mathcal{S}}_{t,n})$;
        \EndIf
        \State \textbf{Return} $\{ \hat{\mathbf{p}}_{t,n} \}_{n = 1}^N$ and $\hat{\boldsymbol{\delta}}_{t}$.
    \end{algorithmic}
\end{algorithm}

\begin{remark}
    Since the ToA set $\mathcal{M} = \{1, 2, \ldots, M\}$ is finite, the enlargement of the total selected ToA subset $\tilde{\mathcal{S}}_{t-1} \subseteq \mathcal{M}$ (line 17 in Algorithm~\ref{alg}) can be executed at most $M$ time instances. 
    This means that when the time instance $t$ is large, the algorithm will predominantly use the reduced BRMP formulas \eqref{eq: RLS_delta} and \eqref{eq:RLS_reduced} (line 13 in Algorithm~\ref{alg}) to update $\hat{\boldsymbol{\delta}}_{t}$.
\end{remark}

\section{Simulation Studies}~\label{sec:simulation}
In this section, we evaluate the proposed algorithm's accuracy in synchronization, NLoS identification, and localization, as well as its computational overhead, through simulations.

For the experiment setup, we consider a square area of $[0, 32] \times [0, 32]\,\text{m}^2$, with $M=25$ anchors positioned at a height of $5$ m and $N = 4$ agents located at a height of $1.5$ m.
The anchor locations $\{ \mathbf{q}_m : m = 1, ..., M\}$ are set as $\{(\frac{32(i-1)}{\sqrt{M}-1}, \frac{32(j-1)}{\sqrt{M}-1}) \, \text{m}:i=1,\cdots,\sqrt{M},\ j=1,\cdots,\sqrt{M}\}$.
At each time instance $t$ ($t = 1, 2, ..., 500$), the agent positions $\{ \mathbf{p}_{t, n}: n = 1, ..., 4 \}$ are randomly generated within the square area using a uniform distribution.
For each agent $n$ at $t$, $12\%$ of the ToA measurements (rounded up) are randomly selected to be affected by severe NLoS errors, i.e., $|\mathcal{S}_{t, n}^{\mathrm{LoS}}| = \lceil0.12M\rceil$, where these NLoS errors $b_{t, n, m}$ follows a uniform distribution over $[10, 40]$ ns.
The clock offset $\delta_m$ is generated from a uniform distribution of $[-8, 8]$ ns.
The STD of measurement noise $\epsilon_{t, n, m}$ is set as $\sigma = 0.4$ ns.
In Algorithm \ref{alg}, we set forgetting factor $\lambda = 0.8$ and proportion parameter $\alpha = 0.88$.
We perform $200$ independent Monte Carlo~(MC) trials. 
The localization performance at each time $t$ is evaluated using the root mean square error (RMSE), defined as 
\begin{align}
    \text{RMSE} = \mathbb{E} \left[ \sqrt{\frac{1}{N}\sum_{n=1}^N \| \hat{\mathbf{p}}_{t,n} - \mathbf{p}^\ast_{t,n} \|^2} \,\right], \label{eq: RMSE}
\end{align}
where $\mathbf{p}^\ast_{t,n}$ is the true position of the $n$-th agent at time $t$.


\begin{figure}[!t]
\centering
\subfloat[]{
		\includegraphics[scale=0.16]{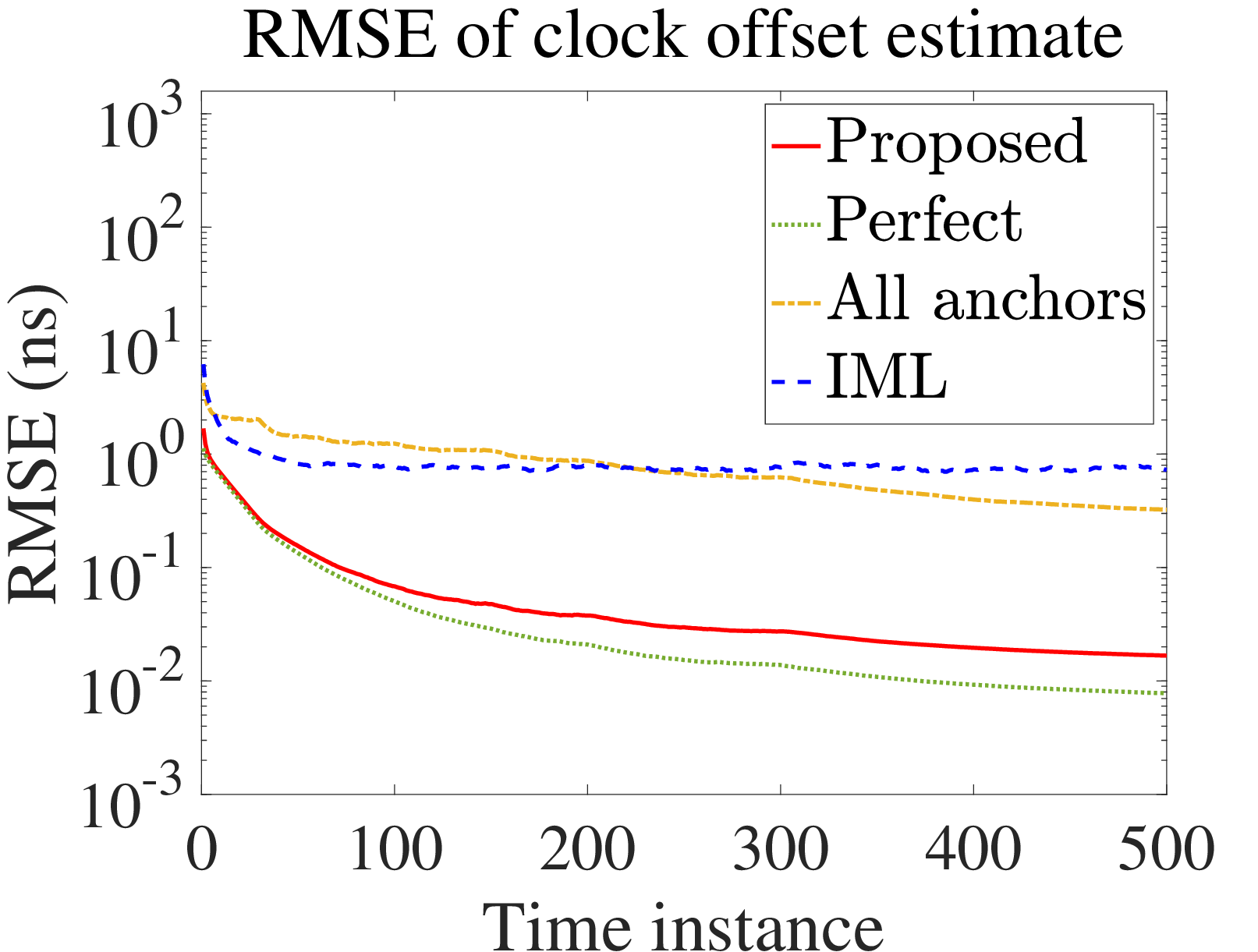}}
\subfloat[]{
		\includegraphics[scale=0.16]{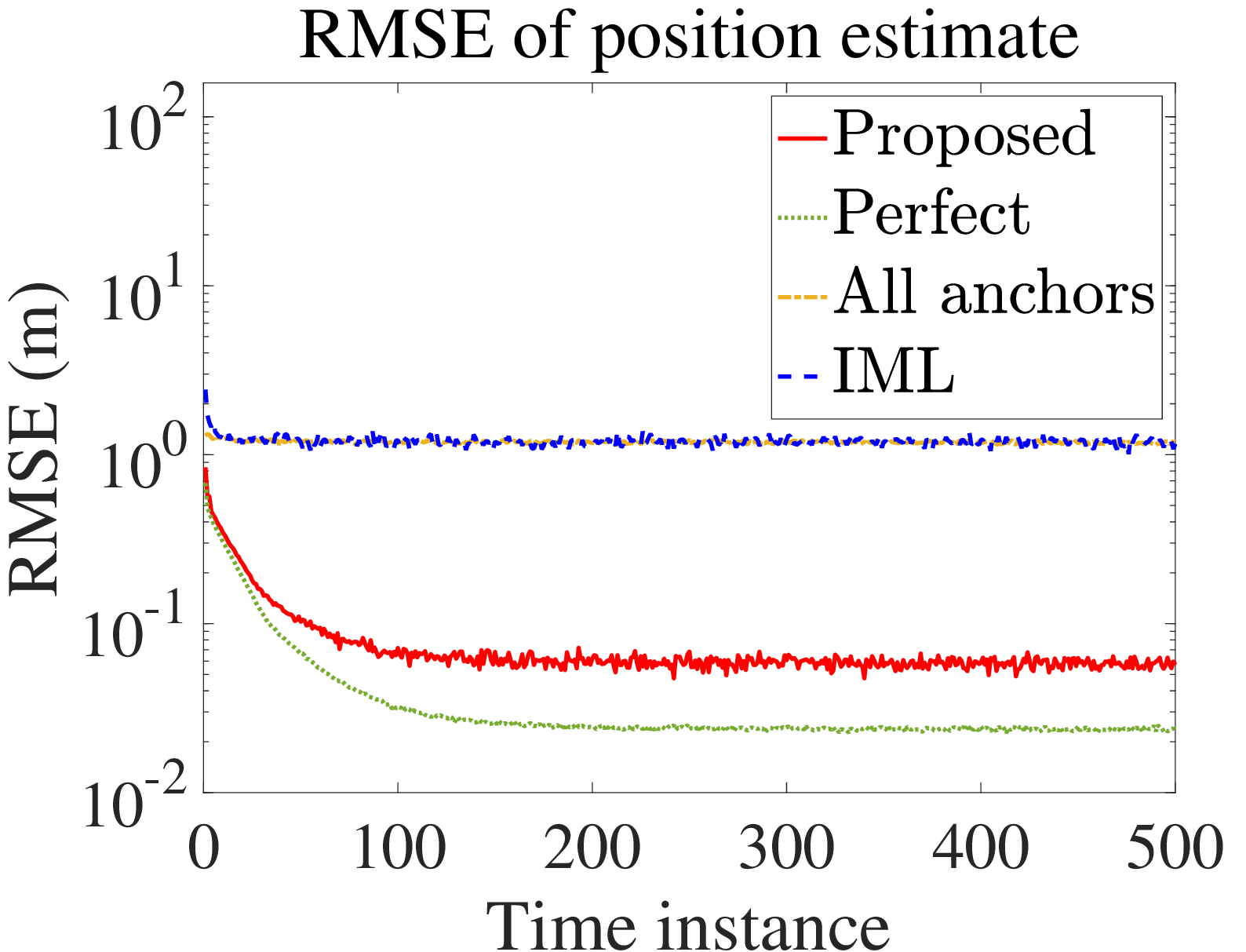}}
        \vspace{-6pt}
    \caption{RMSE of clock offset estimate $\hat{\boldsymbol{\delta}}_{t}$ and agent position estimate $\hat{\mathbf{p}}_{t,n}$ versus time instance $t$ with the number of anchors $M=25$.}
\label{pic: RMSE_time}
\vspace{-0.74cm}
\end{figure}
\begin{figure}[!t]
\centering
\subfloat[]{
		\includegraphics[scale=0.16]{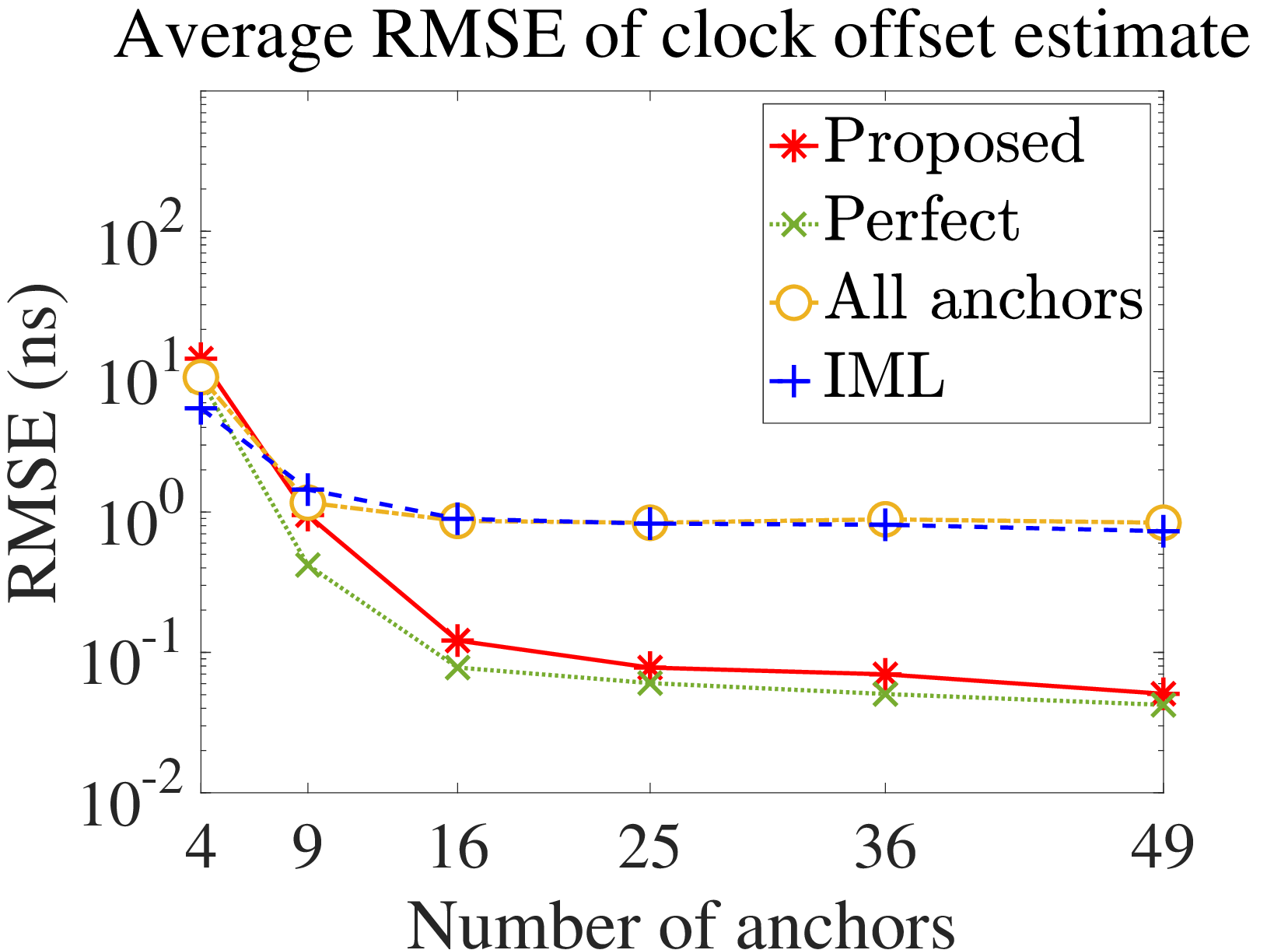}}
\subfloat[]{
		\includegraphics[scale=0.16]{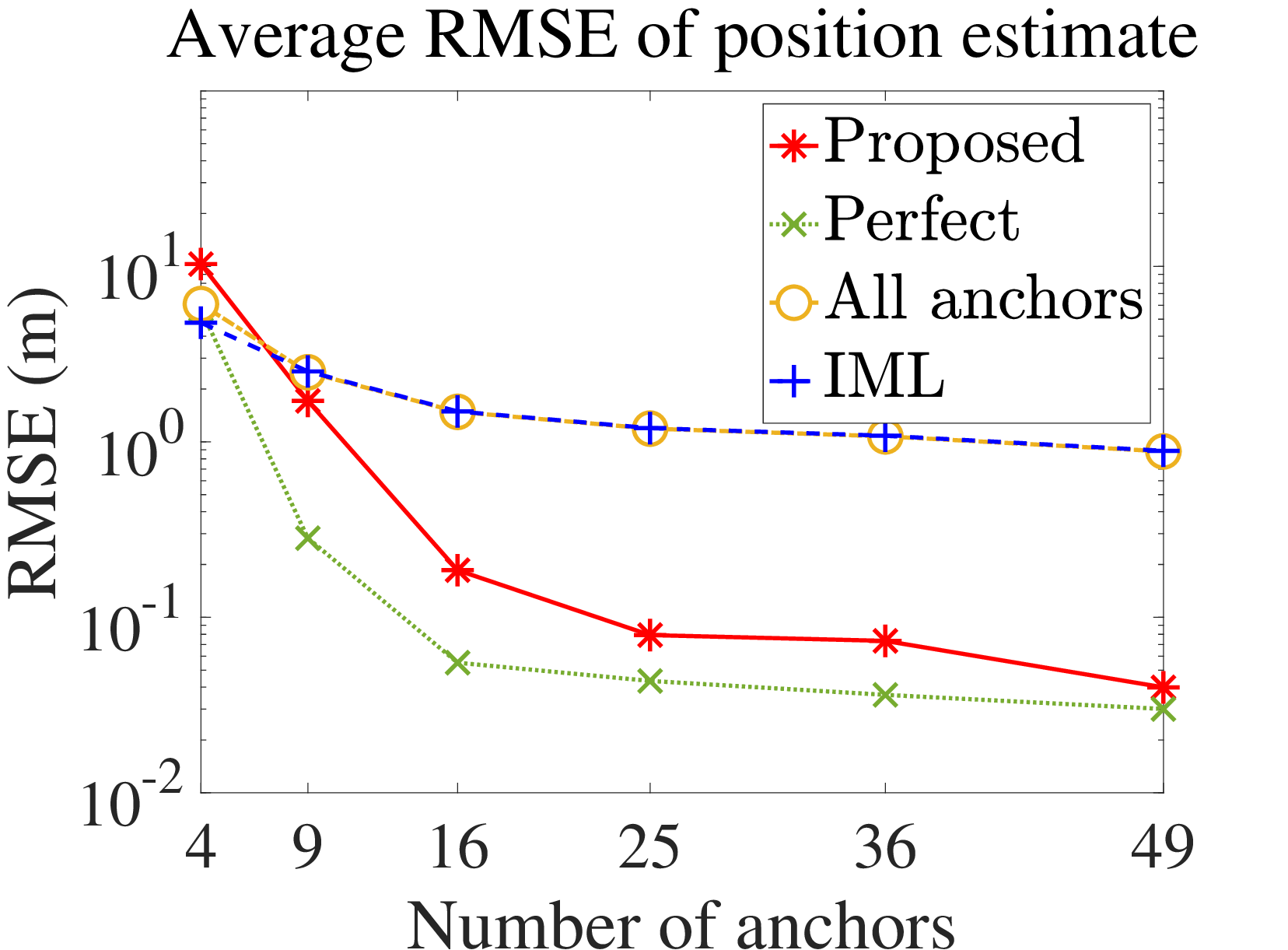}}
        \vspace{-6pt}
    \caption{Average RMSE of clock offset estimate $\hat{\boldsymbol{\delta}}_{t}$ and agent position estimate $\hat{\mathbf{p}}_{t,n}$ across all time instances versus the number of anchors~$M$.}
\label{pic: RMSE_anchor}
\vspace{-0.69cm}
\end{figure}
\begin{figure}[!t]
    \centering
    \subfloat[]{
        \includegraphics[scale=0.16]{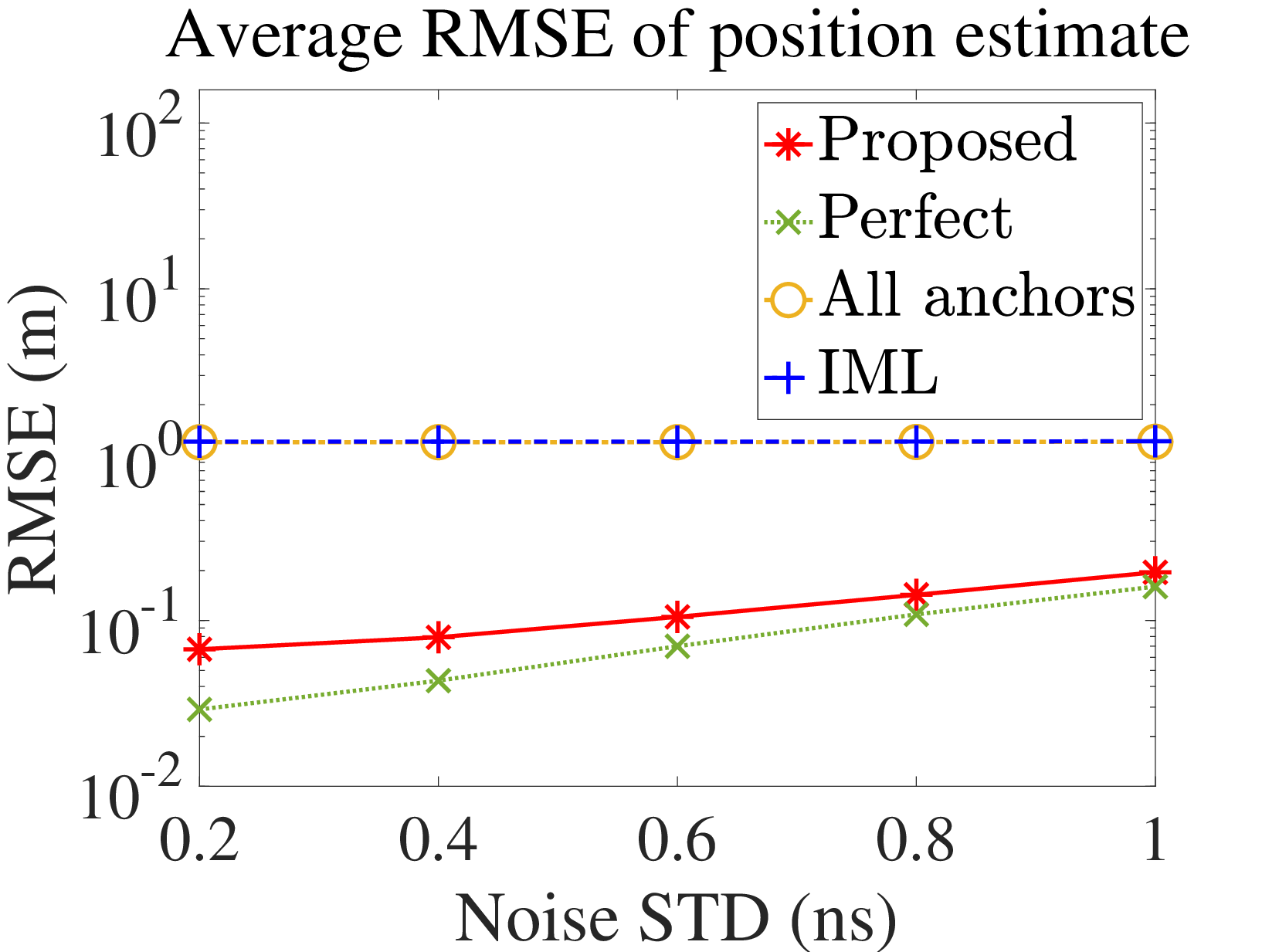}
        \label{pic: sigma}
    }
    \subfloat[]{
        \includegraphics[scale=0.16]{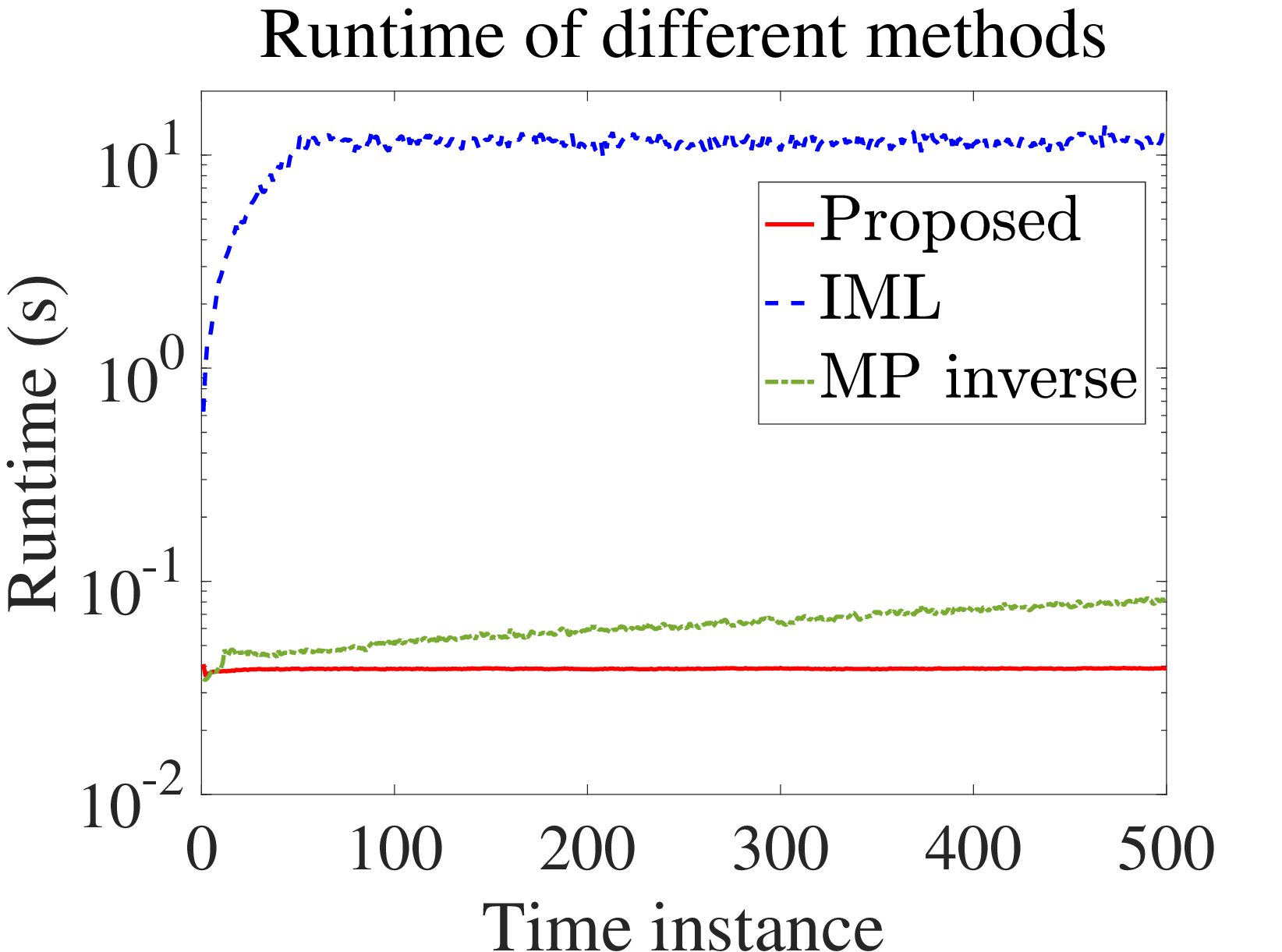}
        \label{pic: time}
    }
    \vspace{-6pt}
    \caption{(a) Average RMSE of agent position estimate $\hat{\mathbf{p}}_{t,n}$ versus noise STD $\sigma$ when $M=25$; (b) Runtime versus time instance $t$.}
    \label{fig:combined}
    \vspace{-0.51cm}
\end{figure}
\begin{table}[t]
    \centering
    \small 
    \caption{NLoS identification accuracy versus noise STD $\sigma$.}
    \label{tab:1}
    \begin{tabular}{m{0.075\textwidth}|m{0.05\textwidth}|m{0.05\textwidth}|m{0.05\textwidth}|m{0.05\textwidth}|m{0.05\textwidth}}
    \Xhline{0.8pt}
    \rule{0pt}{8.5pt}
        $\sigma$ (ns) &  $0.1$ & $0.5$  & $1.0$ & $1.5$ &$2.0$\\
        \hline
        \rule{0pt}{8.5pt}
    Accuracy& 99.55\%	& 99.54\%	&99.30\%	& 95.70\%	 &85.94\%\\
    \Xhline{0.8pt}
    \end{tabular}
    \vspace{-0.61cm}
\end{table}

We compare the synchronization and localization results of the proposed algorithm with two methods.
The first method is the iterative maximum likelihood (IML) algorithm in \cite{jean2014passive}, which performs JLAS using all past ToA measurements.
IML does not include the NLoS identification step and RLS-related techniques for real-time synchronization.
To avoid excessive computational cost, IML is limited to utilizing ToAs from at most 50 prior time instances in the simulations.
The second method is a variant of the proposed algorithm that uses the ToAs from all anchors for localization, i.e., $\hat{\mathcal{S}}^{(k)} = \mathcal{M}$ in \eqref{eq: iter_S}.
Additionally, we introduce a perfect anchor selection method as a benchmark, i.e., $\hat{\mathcal{S}}^{(k)} = \mathcal{S}_{t, n}^{\mathrm{LoS}}$ in \eqref{eq: iter_S}.

Fig.~\ref{pic: RMSE_time} shows the RMSE of clock offset and position estimates versus time instance $t$, respectively.
The RMSE of clock offset estimate is calculated similarly to \eqref{eq: RMSE}.
When $t>100$, the RMSEs of clock offset and position estimates fall below $0.1$ ns and $0.1$ m, respectively. 
Fig.~\ref{pic: RMSE_anchor} shows the average RMSE of clock offset and position estimates across all time instances versus the number of anchors $M$.
It shows that the proposed algorithm achieves superior localization and synchronization accuracy when $M \geq 16$.

Fig.~\ref{fig:combined} (a) illustrates the RMSE of position estimate versus the measurement noise STD $\sigma$. 
TABLE~\ref{tab:1} presents the NLoS identification accuracy, defined as the average percentage of correctly identified NLoS ToAs within the actual NLoS ToA set $\mathcal{S}_{u, n}^{\mathrm{NLoS}}$. Fig.~\ref{fig:combined} (b) illustrates the runtime of different methods versus time instance $t$.
The curve labeled `MP inverse' refers to the direct use of MP inverse in \eqref{eq: Moore-Penrose} to compute $\hat{\boldsymbol{\delta}}_t$.
When $t=500$, the runtime of the MP inverse is $2.05$ times larger than that of BRMP, while both methods achieve the same localization accuracy.
This runtime disparity widens as $t$ increases, owing to the cubic growth in computational and memory overhead of the MP inverse method with respect to~$t$.

\section{Conclusions}~\label{sec:conclusion}
This paper has addressed the challenges of real-time synchronization and localization in large-scale asynchronous WSNs under severe NLoS conditions. 
We have proposed a low-complexity real-time algorithm that introduces the BRMP technique for rapid clock synchronization and employs the RLSR method for NLoS identification and localization.
Simulation results have demonstrated that our algorithm achieves high synchronization and localization accuracy while maintaining low computational overhead, making it suitable for real-time applications in large-scale networks.

\bibliographystyle{IEEEtran}
\bibliography{IEEEabrv, survey_ref}

\begin{thebibliography}{10}
\providecommand{\url}[1]{#1}
\csname url@samestyle\endcsname
\providecommand{\newblock}{\relax}
\providecommand{\bibinfo}[2]{#2}
\providecommand{\BIBentrySTDinterwordspacing}{\spaceskip=0pt\relax}
\providecommand{\BIBentryALTinterwordstretchfactor}{4}
\providecommand{\BIBentryALTinterwordspacing}{\spaceskip=\fontdimen2\font plus
\BIBentryALTinterwordstretchfactor\fontdimen3\font minus \fontdimen4\font\relax}
\providecommand{\BIBforeignlanguage}[2]{{%
\expandafter\ifx\csname l@#1\endcsname\relax
\typeout{** WARNING: IEEEtran.bst: No hyphenation pattern has been}%
\typeout{** loaded for the language `#1'. Using the pattern for}%
\typeout{** the default language instead.}%
\else
\language=\csname l@#1\endcsname
\fi
#2}}
\providecommand{\BIBdecl}{\relax}
\BIBdecl

\bibitem{shastri2022review}
A.~Shastri \emph{et~al.}, ``A review of millimeter wave device-based localization and device-free sensing technologies and applications,'' \emph{{IEEE} Commun. Surveys Tuts.}, vol.~24, no.~3, pp. 1708--1749, May 2022.

\bibitem{laoudias2018survey}
C.~Laoudias \emph{et~al.}, ``A survey of enabling technologies for network localization, tracking, and navigation,'' \emph{{IEEE} Commun. Surveys Tuts.}, vol.~20, no.~4, pp. 3607--3644, Jul. 2018.

\bibitem{zou2020semidefinite}
Y.~Zou and H.~Liu, ``Semidefinite programming methods for alleviating clock synchronization bias and sensor position errors in {TDoA} localization,'' \emph{{IEEE} Signal Process. Lett.}, vol.~27, pp. 241--245, Jan. 2020.

\bibitem{ma2019direct}
F.~Ma, Z.-M. Liu, and F.~Guo, ``Direct position determination in asynchronous sensor networks,'' \emph{{IEEE} Trans. Veh. Technol.}, vol.~68, no.~9, pp. 8790--8803, Jul. 2019.

\bibitem{wang2020tdoa}
T.~Wang, H.~Xiong, H.~Ding, and L.~Zheng, ``{TDoA}-based joint synchronization and localization algorithm for asynchronous wireless sensor networks,'' \emph{{IEEE} Trans. Commun.}, vol.~68, no.~5, pp. 3107--3124, Feb. 2020.

\bibitem{jean2014passive}
O.~Jean and A.~J. Weiss, ``Passive localization and synchronization using arbitrary signals,'' \emph{{IEEE} Trans. Signal Process.}, vol.~62, no.~8, pp. 2143--2150, Feb. 2014.

\bibitem{guo2022new}
N.~Guo \emph{et~al.}, ``New closed-form joint localization and synchronization using sequential one-way {ToAs},'' \emph{{IEEE} Trans. Signal Process.}, vol.~70, pp. 2078--2092, Apr. 2022.

\bibitem{wang2011robust}
Y.~Wang, X.~Ma, and G.~Leus, ``Robust time-based localization for asynchronous networks,'' \emph{{IEEE} Trans. Signal Process.}, vol.~59, no.~9, pp. 4397--4410, Jun. 2011.

\bibitem{yousefi2015sensor}
S.~Yousefi \emph{et~al.}, ``Sensor localization in {NLoS} environments with anchor uncertainty and unknown clock parameters,'' in \emph{IEEE International Conference on Communication Workshop (ICCW)}, London, UK, Jun. 2015, pp. 742--747.

\bibitem{momtaz2018nlos}
A.~A. Momtaz, F.~Behnia, R.~Amiri, and F.~Marvasti, ``{NLoS} identification in range-based source localization: Statistical approach,'' \emph{{IEEE} Sensors J.}, vol.~18, no.~9, pp. 3745--3751, Feb. 2018.

\bibitem{huang2020machine}
C.~Huang \emph{et~al.}, ``Machine learning-enabled {LoS/NLoS} identification for {MIMO} systems in dynamic environments,'' \emph{{IEEE} Trans. Wireless Commun.}, vol.~19, no.~6, pp. 3643--3657, Jan. 2020.

\bibitem{zhuang2021blockwise}
H.~Zhuang, Z.~Lin, and K.-A. Toh, ``Blockwise recursive {Moore}--{Penrose} inverse for network learning,'' \emph{{IEEE} Trans. Syst., Man, Cybern., Syst.}, vol.~52, no.~5, pp. 3237--3250, Mar. 2021.

\bibitem{li2018joint}
S.~Li \emph{et~al.}, ``Joint clock synchronization and position estimation in time of arrival--based passive positioning systems,'' \emph{International Journal of Distributed Sensor Networks}, vol.~14, no.~11, p. 1550147718786893, Nov. 2018.

\bibitem{bhatia2015robust}
K.~Bhatia, P.~Jain, and P.~Kar, ``Robust regression via hard thresholding,'' \emph{Advances in Neural Information Processing Systems}, vol.~28, 2015.

\bibitem{islam2019recursive}
S.~A.~U. Islam and D.~S. Bernstein, ``Recursive least squares for real-time implementation,'' \emph{{IEEE} Control Syst. Mag.}, vol.~39, no.~3, pp. 82--85, May 2019.

\bibitem{ben2003generalized}
A.~Ben-Israel and T.~N. Greville, \emph{Generalized inverses: theory and applications}.\hskip 1em plus 0.5em minus 0.4em\relax Springer Science \& Business Media, 2003, vol.~15.

\bibitem{shanno1970conditioning}
D.~F. Shanno, ``Conditioning of quasi-newton methods for function minimization,'' \emph{Mathematics of Computation}, vol.~24, no. 111, pp. 647--656, Jul. 1970.

\end{thebibliography}

\end{document}